\pdfoutput=1
\documentclass[prl,twocolumn,superscriptaddress,preprintnumbers,nofootinbib]{revtex4}
\usepackage{graphicx}
\usepackage{epsfig}
\usepackage{bm}
\usepackage{latexsym,amssymb,amsmath,amsfonts,amssymb,txfonts,pxfonts,wasysym,float}
\usepackage{mathrsfs}
\usepackage{color}
\usepackage{lettrine}
\usepackage{lipsum}
\usepackage{enumitem}
\usepackage{verbatim}

 \def\be{\begin{equation}}
\def\ee{\end{equation}}
\def\ba{\begin{eqnarray}}
\def\ea{\end{eqnarray}}
\usepackage{pdfsync}

\usepackage[usenames,dvipsnames]{xcolor}
\definecolor{orange}{cmyk}{0,0.5,1,0}
\definecolor{rossoCP3}{cmyk}{0,.88,.77,.40}
\definecolor{graa}{rgb}{0.8,0.8,0.8}
\definecolor{blaa}{rgb}{0.2,0.2,0.6}

\def\simlt{\mathrel{\lower2.5pt\vbox{\lineskip=0pt\baselineskip=0pt
           \hbox{$<$}\hbox{$\sim$}}}}
\def\simgt{\mathrel{\lower2.5pt\vbox{\lineskip=0pt\baselineskip=0pt
           \hbox{$>$}\hbox{$\sim$}}}}

\begin{document}

\preprint{MPP-2021-175}
\preprint{LMU-ASC 42/21}

\title{\color{rossoCP3} Leptophilic $\bm{U(1)}$ Massive Vector Bosons
  from Large Extra Dimensions:\\ Reexamination of Constraints from LEP
Data}

\author{Luis A. Anchordoqui}

\affiliation{Department of Physics and Astronomy,  Lehman College, City University of
  New York, NY 10468, USA}

\affiliation{Department of Physics,
 Graduate Center, City University
  of New York,  NY 10016, USA}

\affiliation{Department of Astrophysics, 
 American Museum of Natural History, NY
 10024, USA}

\author{Ignatios Antoniadis}

\affiliation{Laboratoire de Physique Th\'eorique et Hautes \'Energies - LPTHE
Sorbonne Universit\'e, CNRS, 4 Place Jussieu, 75005 Paris, France}


\author{Xing~Huang}
\affiliation{Institute of Modern Physics, Northwest University, Xi'an
  710069, China}
\affiliation{NSFC-SPTP Peng Huanwu Center for Fundamental Theory,
  Xi'an 710127, China}
\affiliation{Shaanxi Key Laboratory for Theoretical Physics Frontiers,
  Xi'an 710069, China}

\author{Dieter L\"ust}

\affiliation{Max--Planck--Institut f\"ur Physik, 
 Werner--Heisenberg--Institut,
80805 M\"unchen, Germany
}

\affiliation{Arnold Sommerfeld Center for Theoretical Physics 
Ludwig-Maximilians-Universit\"at M\"unchen,
80333 M\"unchen, Germany
}

\author{Fran\c cois Rondeau}

\affiliation{Laboratoire de Physique Th\'eorique et Hautes \'Energies - LPTHE
Sorbonne Universit\'e, CNRS, 4 Place Jussieu, 75005 Paris, France}

\author{Tomasz R. Taylor}

\affiliation{Department of Physics,
  Northeastern University, Boston, MA 02115, USA}

\begin{abstract}
  \vskip 2mm \noindent Very recently, we proposed an explanation of
  the discrepancy between the measured anomalous magnetic moment of
  the muon and
  the Standard Model (SM) prediction in which the dominant contribution to $(g-2)_\mu$
  originates in Kaluza-Klein (KK) excitations (of the lepton gauge
  boson) which do not mix with quarks (to lowest order) and therefore
  can be quite light avoiding LHC constraints. In this addendum we
  reexamine the bounds on 4-fermion contact interactions from precise
  electroweak measurements and show that the constraints on KK masses
  and couplings are more severe than earlier thought. However, we
  demonstrate that our explanation remains plausible if a few KK modes
  are lighter than LEP energy, because if this were the case the
  contribution to the 4-fermion scattering from the internal
  propagator would be dominated by
  the energy and not by the mass. To accommodate the $(g-2)_\mu$
  discrepancy we assume that the lepton number $L$ does not partake in
  the hypercharge and propagates in one extra dimension (transverse to
  the SM branes): for a mass of the lowest KK excitation of 60~GeV (lower
  than the LEP energy), the string scale is roughly 10~TeV while the
  $L$ gauge coupling is of order $\sim 10^{-1}$.
\end{abstract}

\maketitle

In~\cite{Anchordoqui:2021vrg} we argue that the exchange of Kaluza-Klein (KK) excitations of the lepton number ($L$) gauge boson could provide a dominant contribution to $(g-2)_\mu$ and explain the discrepancy between the Standard Model (SM) prediction of $a_\mu = (g-2)_\mu/2$ and experiment: $\Delta a_\mu^{\rm exp} \equiv a_\mu^{\rm FNAL + BNL} - a_\mu^{\rm SM} = (251 \pm 59) \times 10^{-11}$~\cite{Abi:2021gix}. On the other hand, the zero mode of the lepton number gauge boson is anomalous and gains a mass ${\cal O}(M_s)$ through a four-dimensional generalisation of the Green-Schwarz anomaly cancellation mechanism. Its mass being at the string scale, its contribution to $(g-2)_\mu$ is negligible, and therefore only the contributions of the KK modes are relevant to explain the discrepancy. In this addendum we reexamine model constraints from LEP data. 

At the leading  order in the $U(1)_L$ coupling constant $g_L$, the contribution of massive vector bosons to $(g-2)_\mu$ comes from the muon vertex correction, and is given by
\begin{equation}\label{eq:integral}
\Delta a_{\mu}=\frac{\alpha_L m_{\mu}^2}{\pi}\int_0^1dx dy dz \delta(x+y+z-1) \frac{z(1-z)}{(1-z)^2m_{\mu}^2+zM^2},
\end{equation}
where $M$ is the mass of the boson, $m_{\mu}$ the muon mass and $\alpha_L = g_L^2/(4\pi)$. One can then consider three different cases, depending whether $M \gg m_{\mu}$, $M\sim m_{\mu}$ or $M\ll m_{\mu}$. 

\subsection{Case 1 : $M \gg m_{\mu}$}
When all KK states have masses much bigger than the muon mass, the sum of the integral \eqref{eq:integral} over all the KK states can be approximated by
\begin{equation}\label{uno}
\Delta a_\mu^{(1)}= \sum_n\frac{1}{3}{\alpha_L(n)\over \pi}{m_{\mu}^2\over M_n^2},
\end{equation}
where $M_n$ is the mass of the $n$th KK excitation~\cite{Anchordoqui:2021vrg}.

The bound from LEP data on the so-called compositeness scale associated to 4-fermion operators is given by~\cite{ParticleDataGroup:2020ssz}:
\begin{equation}
\left|\sum_n \frac{\alpha_L(n)}{s-M_n^2}\right| < B \sim (10~{\rm TeV})^{-2} \,,
\label{dos}
\end{equation}
where $s$ is the square of the center-of-mass energy\footnote{For fine-tuned values of $M_n$ close to $\sqrt{s}$, the vector boson propagator appearing in the left-hand side of \eqref{dos} is regulated by replacing $\frac{1}{s-M_n^2}$ by $\frac{1}{s-M_n^2+i\Gamma_nM_n}$, with $\Gamma_n$ the decay rate of the $n$-th KK mode. Since the number of possible decay channels of the KK excitations increase for higher modes, $\Gamma_n$ increases with $n$ and its explicit computation would require a model dependent analysis.}. For $M_n \gg \sqrt{s}$, (\ref{dos}) reduces to $\sum_n \alpha_L(n)/M_n^2 < B$. Thus, the sum of the KK exchange given in (\ref{uno}) is constrained by the compositeness bound, yielding $\Delta a_\mu^{(1)} \sim {\cal O}(10^{-11})$; a result which is independent on the number of extra dimensions. Hence, one needs at least few KK modes lighter than LEP energy in order to provide a significant contribution able to bridge the gap in the muon anomalous magnetic moment.

A crucial point to take into account is that the gauge coupling is suppressed by the volume of the compact space $V_\perp \sim (R M_s)^d$,
\begin{equation}
  g^2_L = g_s/V_\perp,
\end{equation}
where $g_s$ is the string coupling, $R$ is the compactification scale, $M_s$ is the string scale, and $d$ stands for the number of extra dimensions in which $L$ propagates. For $d=1$, we have $M_n = n/R$ and after substituting these figures into (\ref{uno}), $\Delta a_\mu^{(1)}$ becomes \footnote{We have neglected here the $n$-dependence of the gauge coupling of the $n$-th KK excitation, given in the case of one extra dimension by $g_L(n)=g_L\exp\left\{-cn^2\frac{M_1^2}{M_s^2}\right\}$, with $c$ a positive (model dependent) numerical constant. When $M_1\ll M_s$, as it is the case in the large extra dimension scenario considered in this letter, the exponential is of order $1$ for all $n\simlt\frac{M_s}{M_1}$, and the gauge coupling can indeed be taken constant. The exponential suppression of $g_L$ becomes significant only for higher KK modes with $n\gg\frac{M_s}{M_1}$, which give a negligible contribution to $\Delta a_{\mu}^{(1)}$.}
\begin{equation}
\Delta a_\mu^{(1)}=\frac{g_sm_{\mu}^2}{72M_1M_s}.
\end{equation}
The observed value of $\Delta a_\mu$ then implies
\begin{equation}\label{eq:M1_Ms_relation}
M_1 \ M_s \sim g_s\times 5 \times 10^4~{\rm GeV}^2,
\end{equation}
where $g_s \simlt 4\pi$ to remain in the perturbative regime.

As an illustration, if we take $M_s = 10~{\rm TeV}$ then we have $M_1 \sim g_s\times 5~{\rm GeV}$, so that the highest possible value for the compactification scale $M_1$, obtained for $g_s=4\pi$, is of order $M_1\sim 60~{\rm GeV}$, which is consistent with the condition $m_{\mu} \ll M_1 \ll \sqrt s$ for all the approximations. The associated gauge coupling is then of order $g_L\sim 10^{-1}$. Taking $\left. \sqrt{s}\right|_{\rm LEP}=209~{\rm GeV}$, the total KK contribution to the LEP bound is given by
\begin{equation}
\left|\sum_n \frac {g_L^2}{4 \pi (s-n^2M_1^2)}\right|\sim 10^{-2}~{\rm TeV}^{-2},
\end{equation}
and hence the bound (\ref{dos}) is satisfied.

We also note that we have used a bound on $g_L$ for masses lighter than the LEP center-of-mass energy (by neglecting the mass compared to the energy in the exchange $Z'$ propagator) using the bound on new physics compatible with the bound on the compositeness scale.

Note that to lower the string scale in the region discussed above, one assumes in general additional large extra dimensions transverse to both SM and $L$ stacks of branes that do not play any role in our analysis.

\subsection{Case 2 : $M \sim m_{\mu}$}
In the case of a massive boson with a mass of order of the muon mass $m_{\mu}$, its contribution \eqref{eq:integral} to $(g-2)_{\mu}$ is given by 
\begin{equation}\label{eq:contrib_M=m_mu}
\Delta a_{\mu}^{(2)}=\frac{\alpha_L}{\pi}\frac{-9+2\sqrt{3}\pi}{18}.
\end{equation}
If the lightest KK state have a mass $M_1\sim m_{\mu}$, the total contribution to the muon anomalous magnetic moment is therefore the sum of $\Delta a_{\mu}^{(1)}$ (Eq. \eqref{uno} for $n>1$) and $\Delta a_{\mu}^{(2)}$ (Eq. \eqref{eq:contrib_M=m_mu}), which in the case of one extra dimension yields:
\begin{equation}
\Delta a_{\mu}=\frac{g_s}{4\pi^2}\frac{m_{\mu}}{M_s}\left(\frac{-9+2\sqrt{3}\pi}{18}+\frac{1}{3}\sum_{n>1}\frac{1}{n^2}\right).
\end{equation} 
The $(g-2)_{\mu}$ discrepancy can then be accommodated for a string scale at $M_s\sim g_s\times 3 \times 10^2~{\rm TeV}$, yielding a coupling $g_L\sim 5\times 10^{-4}$, now independent of $g_s$. With $M_1=m_{\mu}=105~{\rm MeV}$, we now get
\begin{equation}
\left|\sum_n \frac {g_L^2}{4 \pi (s-n^2M_1^2)}\right| \sim 10^{-4}~{\rm TeV}^{-2},
\end{equation}
so that the bound (\ref{dos}) is also satisfied.

\subsection{Case 3 : $M \ll m_{\mu}$}
We can also consider the situation where some of the lightest KK states have masses much lower than the muon mass, in which case the integral \eqref{eq:integral} gives a constant contribution $\frac{\alpha_L}{2\pi}$. Multiplying by $\frac{m_{\mu}}{M_1}$, the number of states with masses below $m_{\mu}$, and assuming again one extra dimension, we get the contribution
\begin{equation}\label{eq:contrib_M<m_mu}
\Delta a_{\mu}^{(3)} = \frac{g_s}{8\pi^2}\frac{m_{\mu}}{M_s}.
\end{equation}
The total contribution to the muon anomalous magnetic moment is then the sum of $\Delta a_{\mu}^{(1)}$ (Eq. \eqref{uno} for $n>\frac{m_{\mu}}{M_1}+1$), $\Delta a_{\mu}^{(2)}$ (Eq. \eqref{eq:contrib_M=m_mu}) and $\Delta a_{\mu}^{(3)}$ (Eq. \eqref{eq:contrib_M<m_mu}), that is, in the case of one extra dimension:
\begin{equation}
\Delta a_{\mu}=\frac{g_s}{8\pi^2}\frac{m_{\mu}}{M_s}\left(1+2\cdot\frac{-9+2\sqrt{3}\pi}{18}+\frac{2}{3}\frac{m_{\mu}}{M_1}\sum_{n=\frac{m_{\mu}}{M_1}+2}\frac{1}{n^2} \right).
\end{equation}
As an example, let us take $\frac{m_{\mu}}{M_1}=10$, in which case $\Delta a_{\mu}\sim\frac{g_s}{8\pi^2}\frac{m_{\mu}}{M_s}$, accommodating the discrepancy for a string scale $M_s\sim g_s\times 5 \times 10^2~{\rm TeV}$. With $M_1=\frac{m_{\mu}}{10}=10.5~{\rm MeV}$, one gets a coupling $g_L\sim 10^{-4}$, again independent of $g_s$, from which we can evaluate
\begin{equation}
\left|\sum_n \frac {g_L^2}{4 \pi (s-n^2M_1^2)}\right| \sim 6\times 10^{-4}~{\rm TeV}^{-2},
\end{equation}
again satisfying the bound (\ref{dos}).\\

Let us note that unlike the discrepancy between the experimental value and the SM prediction of the muon anomalous magnetic moment which is positive, $\Delta a_\mu^{\rm exp} \equiv a_\mu^{\rm exp} - a_\mu^{\rm SM} = (251 \pm 59) \times 10^{-11}$, the discrepancy of the electron anomalous magnetic moment is negative, $\Delta a_e^{\rm exp} \equiv a_e^{\rm exp} - a_e^{\rm SM} = - 88(36) \times 10^{-14}$~\cite{Aoyama:2019ryr}. The contributions coming from the KK excitations being positive, they will increase the discrepancy of $(g-2)_e$, and we thus have to check that this contribution is lower than or of order of the experimental error on $(g-2)_e$, that is $\simlt 10^{-13}$. Assuming $M_1\gg m_e$ where $m_e$ is the electron mass, this contribution is simply obtained by replacing the muon mass $m_{\mu}$ by the electron mass $m_e$ in \eqref{uno}, namely
\begin{equation}
\Delta a_e = \frac{m_e^2}{m_{\mu}^2} \Delta a_{\mu}^{(1)}=\frac{m_e^2 g_s}{72M_1M_s}.
\end{equation}
For the different values obtained above for $M_1$ and $M_s$, we get in the case 1 $\Delta a_e\sim 10^{-14}$, and in the cases 2 and 3 $\Delta a_e\sim 10^{-13}$, indeed smaller than or of order of the error on $(g-2)_e$.\\

Finally, one may also worry about LHC bounds using the one loop lepton induced mixing between the $L$ KK-excitations and the photon or $Z$. The latter couples to quarks while the former can couple to a dilepton pair. The corresponding Drell-Yan exchange can then be estimated as:
\begin{equation}
{1\over E^2}g_L^2\times N\times 10^{-2} \simeq {10^{-2}\over EM_s}< (5\,{\rm TeV})^{-2}
\end{equation}
where $E$ is the dilepton energy, $N\simeq E/M_1$ is the number of
KK-modes with mass less than $E$, $g_L^2\simeq g_s M_1/M_s$ and $10^{-2}$ counts for the loop factor suppression. It follows that the proposed scenario is compatible with LHC bounds~\cite{Greg}.\\ 

We thank Greg Landsberg for valuable discussion. The work of L.A.A. is supported by the 
  U.S. National Science Foundation (NSF Grant PHY-2112527). The research of I.A. was partially performed as
  International professor of the Francqui Foundation, Belgium. The
  work of X.H. is supported by the NSFC Grant No. 12047502. The
  work of D.L. is supported by the Origins Excellence Cluster. The
  work of T.R.T is supported by NSF under Grant Number PHY-1913328.
  Any opinions, findings, and conclusions or recommendations expressed
  in this material are those of the authors and do not necessarily
  reflect the views of the NSF.

\end{document}